\magnification = \magstep1
\baselineskip = 24 true pt
\hsize = 16 true cm
\vsize = 22 true cm
\def\dddot#1{\mathord{\mathop{\kern0pt#1}\limits^{%
  \vbox{\hbox{.\kern-.075em.\kern-.075em.\kern-.025em}\vskip-.15ex}}}}
\centerline{\bf THE MAXIMAL INVARIANCE GROUP OF NEWTON'S } 
\centerline{\bf EQUATIONS FOR A FREE POINT PARTICLE$^{\dag}$} 
\bigskip
\medskip
\centerline {O. Jahn and V. V. Sreedhar}
\centerline {School of Theoretical Physics}
\centerline {Dublin Institute for Advanced Studies}
\centerline {10, Burlington Road}
\centerline {Dublin 4, Ireland}
\bigskip
\centerline {\bf Abstract}

The maximal invariance group of Newton's equations for a free nonrelativistic
point particle is shown to be larger than the Galilei group. It is a 
semi-direct product of the static (nine-parameter) Galilei group and an 
$SL(2,R)$ group containing time-translations, dilations and a one-parameter 
group of time-dependent scalings called {\it expansions}. This group was first 
discovered by Niederer in the context of the free Schr\"odinger equation. We 
also provide a road map from the free nonrelativistic point particle to the 
equations of fluid mechanics to which the symmetry carries over. The hitherto 
unnoticed $SL(2, R)$ part of the symmetry group for fluid mechanics gives a 
theoretical explanation for an observed similarity between numerical 
simulations of supernova explosions and numerical simulations of experiments 
involving laser-induced implosions in inertial confinement plasmas. We also 
give examples of interacting many body systems of point particles which have 
this symmetry group.  
$$ $$
\noindent $~^{\dag}${\it Dedicated~to~the~memory~of~Prof.~L.~O'Raifeartaigh}
\hfil \break
\vfill
\hfill DIAS-STP-01-02\hfil\break
\vfil\eject
Almost all introductory books on the special theory of relativity mention, at 
least in passing, that Newton's equations of motion for a classical free 
nonrelativistic point particle are invariant under Galilei transformations. 
Probably not many eyebrows would be raised if we jumped from this fact to the 
conclusion that the Galilei transformations are the most general coordinate 
changes under which Newton's equations retain their form. It would therefore 
come as a considerable surprise to learn that there are other transformations 
which do the same. A simple example of such transformations may be given by 
noting that a freely moving point particle, with initial position ${\bf x}_0$ 
and velocity ${\bf v}_0$, traverses a straight line ${\bf x}(t ) = {\bf x}_0 + 
{\bf v}_0t$. This equation may be rewritten as ${\bf x}/t = {\bf v}_0 + 
{\bf x}_0/t$. In this representation $t$ is replaced by $1/t$, lengths are 
scaled by a factor proportional to time, and the initial position and velocity 
are interchanged; but the important point is that the trajectory remains a 
straight line. Hence, the new variables ${\bf x}/t,~1/t$ satisfy the same 
equations as the old ones ${\bf x},~t$ [1]. Therefore the following question 
naturally arises: What is the maximal invariance group of Newton's equations of
motion for a classical free nonrelativistic point particle? The answer to this 
question reveals that the maximal invariance group is a twelve-parameter group 
consisting of the usual ten-parameter Galilei group, the one-parameter group of
dilations, and a one-parameter group of time-dependent scalings called 
{\it expansions} which are nonrelativistic analogues of special conformal 
transformations. The existence of these transformations is not merely of 
academic interest. As explained in [1, 2], such transformations provide a 
theoretical explanation for the plausibility of simulating astrophysical 
systems like supernova explosions by performing laser-induced plasma 
implosions. 

In order to find the maximal invariance group of an equation, one has to 
find the set of all spacetime transformations which leave it form-invariant
up to a factor [3]. This condition is equivalent to the requirement that the 
Action be invariant. The Action $S$ for a free point particle of mass $m$, 
in $d$ space dimensions, at position $x_i(t)$ is given by 
\vfill\eject
$$S = {m\over 2}\int dt~\bigl({dx_i\over dt}\bigr)^2 \eqno(1)$$ 
where $i ~=~ 1\cdots d$ and the sum over the index $i$ has been suppressed.
Let us now consider a transformation to new coordinates $\xi,~\tau$  
$$\xi_i =  f_i(x, t),~~~~~~~\tau = h(x, t)\eqno(2)$$
We wish to find the most general functions $f_i,~h$ which leave the Action
(1)  form-invariant:  
$$\int dt~\bigl({dx_i\over dt}\bigr)^2 
= \int d\tau~\bigl({d\xi_i\over d\tau}\bigr)^2\eqno(3)$$ 
This is achieved by requiring
$${\Bigl({\partial f_i\over\partial x_j}{dx_j\over dt} + {\partial f_i
\over\partial t}\Bigr)^2\over\Bigl({\partial h\over\partial x_k}{dx_k\over dt} 
+ {\partial h\over\partial t}\Bigr)}~~~ = ~~~\bigl({dx_i\over dt}\bigr)^2 + 
{d\over dt} F(x, t)\eqno(4)$$ 
for arbitrary functions $x_i(t)$ where $F(x, t)$ is an arbitrary boundary term
and 
$${d\over dt}F(x, t) =  \Bigl({\partial F\over\partial x_i}{dx_i\over dt} + 
{\partial F\over\partial t}\Bigr) \eqno(5)$$
Comparing powers of $dx_i/dt$ on both sides of (4), we get 
$$ {\partial h\over \partial x_i} = 0~~~ ~~~\Rightarrow~~~~~~ \tau = h(t)
\eqno(6)$$
$$\bigl({\partial f_i\over\delta x_j}\bigr)
\bigl({\partial f_i\over\delta x_k}\bigr)
~~~ = ~~~{\partial h\over \partial t}\delta_{jk}\eqno(7)$$ 
$$ 2{\partial f_i\over\partial x_j}{\partial f_i\over \partial t}
~~~=~~~ {\partial F\over \partial x_j}{\partial h\over \partial t}\eqno(8)$$ 
$$ \bigl({\partial f_i\over\partial t}\bigr)^2
~~~=~~~ {\partial F\over\partial t}{\partial h\over\partial t}\eqno(9)$$ 
From (6) and (7) it follows that 
$$ T_{ljk} + T_{lkj} = 0~~~~{\hbox{where}}~~~T_{ljk} = {\partial^2 f_i\over
\partial x_l\partial x_j}{\partial f_i\over\partial x_k}\eqno(10)$$
Subtracting $T_{kjl} + T_{klj}$ from the first equation in (10) and using the 
symmetry of $T_{ljk}$ under a permutation of the first two indices, it may be 
shown that $T_{ljk}$ is a totally symmetric tensor. Therefore $T_{ijk} = 0$, 
which implies that $f_i$ is linear in $x$,
$$f_i(x, t)~~ =~~ l(t)R_{ij} (t)x_j  + m_i (t) \eqno (11)$$
It follows from (7) that $R$ can be chosen to be orthogonal, $R^T(t)R(t) = 1$. 
Further, differentiating (8) with respect to $x_k$ and using the explicit 
expression for $f$ from (11) we get 
$$ (l\dot l) \delta_{kj} + l^2 (R^T\dot R)_{kj}~~~ = ~~~{1\over 2}{\partial F
\over \partial x_k\partial x_j} \dot h\eqno(12) $$ 
where the dot refers to a derivative with respect to $t$. Since the right hand 
side and the first term on the left hand side are symmetric in $k$ and $j$,  
whereas the second term on the left hand side is antisymmetric, by virtue of 
being in the Lie algebra of the rotation group; we have $\dot R = 0$ and thus, 
$R$ is a constant (rigid) rotation matrix. It then follows from (7) that 
$$\dot h = l^2\eqno(13)$$
Eliminating $F$ from (8) and (9) we get 
$${\partial^2 f_i\over \partial t^2}{\partial h\over \partial t}
~~~=~~~{\partial f_i\over\partial t}{\partial^2 h\over\partial t^2}\eqno(14)$$ 
Substituting (11) into this equation and  comparing powers of $x$ we find
$$l\ddot l = 2(\dot l)^2 \eqno(15)$$
$$l\ddot m_i = 2(\dot m_i\dot l) \eqno(16)$$
Using (13), equation (15) can be rewritten in terms of $h(t)$ and takes the 
form  
$${\dddot h\over\dot h} - {3\over 2}\Bigl({\ddot h\over\dot h}\Bigr)^2 
= 0 \eqno(17)$$ 
The left hand side of the above equation is called the Schwarzian derivative of
$h$ and a standard result of complex analysis -- although only the real part 
is relevant here -- is that the solution of the above differential equation is 
$$h (t) = {\alpha t + \beta\over\gamma t + \delta}~~~~~~{\hbox{where}}~~~~~~
\alpha \delta - \beta\gamma = 1\eqno(18)$$
where $\alpha,~\beta,~ \gamma,~\delta$ are real. These transformations go by 
various names: fractional linear, projective, and  global conformal. They form 
the group $SL(2, R)$. Substituting the above result in (13), and solving (16) 
for $m_i(t)$ we get 
$$ l(t) = {1\over \gamma t + \delta},~~~~~~~~m_i(t) = {b_i\over
\gamma t + \delta} + d_i\eqno(19)$$
where $b_i$ and $d_i$ are integration constants. From (11) and (18) we then 
have, for the most general transformations that leave Newton's equations 
invariant, 
$$\xi_i = {R_{ij}x_j + a_i + v_it\over \gamma t + \delta},~~~~~~\tau = 
{\alpha t + \beta\over\gamma t + \delta}~~~~ {\hbox{where}}~~~~\alpha \delta - 
\beta\gamma = 1,~~~R^TR = 1\eqno(20)$$
where $a_i$ and $v_i$ are constants expressible in terms of $b_i$ and $d_i$.  
It is useful to consider the following two special cases:  
\bigskip
\noindent {\it I. $\beta = \gamma = 0,~\alpha~ =~ \delta~ =~ 1$: Connected,
Static Galilei Group $G$:}
In this case, we have 
$$g:~~~\tau = t,~~~~~{\bf \xi} = R{\bf x + a + v}t \eqno(21)$$
These equations describe connected, static Galilei transformations which 
exclude parity and time-reversal. It is clear from (21) that this is a
nine-parameter group.  
\medskip
\noindent{\it II. ${\bf a = v = 0},~ R = 1:~ SL(2, R)$ Transformations:} 
In this case, we have 
$$\sigma:~~~~~\tau = {\alpha t + \beta\over\gamma t + \delta},\qquad {\bf\xi} = 
{{\bf x}\over\gamma t + \delta};~~~~\alpha\delta - \beta\gamma = 1\eqno(22) $$
These are the $SL(2,R)$ generalisations of the inversion transformations 
presented in [1] and include time translations ($\gamma = 0,~\alpha
=\delta = 1$), dilations ($\beta = \gamma = 0$) and a one-parameter group 
of time-dependent scalings called {\it expansions} ($\alpha = \delta = 1,~
\beta = 0$). Since the parameters are constrained by the condition 
$\alpha\delta - \beta\gamma = 1$, $SL(2, R)$ is a three-parameter group. 
\bigskip
To understand the structure of the group, we study the relationship between the
$SL(2, R)$ group and the connected static Galilei group $G$. Let us first 
consider a conjugation of a $g\in G$ by a $\sigma\in SL(2, R)$. By making 
three successive transformations of $x$ and $t$ we find that 
$$\sigma^{-1}(\alpha , \beta , \gamma ,\delta )g(R, {\bf a}, {\bf v}) \sigma 
(\alpha , \beta , \gamma ,\delta ) = g(R, {\bf a_\sigma,  v_\sigma})\eqno(23)$$
where 
$$\pmatrix{{\bf v_\sigma}\cr {\bf a_\sigma} } = \pmatrix {\alpha & \gamma\cr
\beta & \delta } \pmatrix{{\bf v}\cr {\bf a}}\eqno(24)$$
which shows that $G$ is an invariant subgroup. This result can be used to 
determine the product of two general elements $\sigma g$ and $\sigma'g'$ of 
the full group,
$$\sigma g \sigma' g' = \sigma \sigma' {\tilde g} g',~~~{\hbox{where}}~~~
\tilde g = \sigma'^{-1} g \sigma'\in G \eqno (25) $$ 
This shows that the full group is not a direct, but only a semi-direct product
$$ {\cal G} = SL(2,R) \wedge G   \eqno (26) $$ 
As is apparent from (25), the two factors in the semi-direct product are on a 
different footing: while $G$ is an invariant subgroup, $SL(2,R)$ is not. 
Furthermore, recall that $G$ itself takes the form
$$ G = R \wedge ( T \otimes B )\eqno(27) $$
where $R$ is the rotation group and $T$ and $B$ are translation and boost 
groups with parameters ${\bf a}$ and ${\bf v}$ respectively. Since $SL(2,R)$ 
commutes with $R$, ${\cal G}$ can be expressed as a single semi-direct product
$${\cal G} = \bigl( SL(2,R)\otimes R\bigr)\wedge ( T\otimes B)\eqno(28) $$
Further it may be noted that the inversion $\Sigma$ considered in [1] is 
the special element of $SL(2, R)$ for which $(\alpha ,~\beta ,~\gamma ,~\delta
 ) = (0,~ -1,~ 1,~ 0)$. Note that $\Sigma^2 = P$ where $P$ is the parity
transformation. This observation can be used to give a novel interpretation to 
a Galilei transformation. To see this we consider the coset elements 
$g_{\atop\Sigma}(R,~ {\bf a},~ {\bf v})\equiv\Sigma g(R,~{\bf a},~{\bf v})$, 
where $g \in G$, we have   
$$\eqalign{g_{\atop\Sigma} (R',~ {\bf a}',~ {\bf v}')g_{\atop\Sigma}(R,~{\bf 
a},~{\bf v}) &= g_{\atop P}(R'R,~R'{\bf a -  v}',~R'{\bf v + a}')\cr
\Rightarrow~~~ g_{\atop \Sigma}^2(R,~{\bf a,~v}) &= g_{\atop P}(R^2, R{\bf a -
 v}, R{\bf v +  a})}\eqno(29)$$
where we have used the obvious notation $g_{\atop P} (R, ~{\bf a, ~v}) = Pg
(R,~{\bf a, v})$. Since every pair of vectors can be expressed as 
$R{\bf a - v}$ and $R{\bf v+a}$ for suitable {\bf a} and {\bf v}, this shows 
that every parity reflected static Galilei transformation is the square of a 
coset transformation $g_{~\atop\Sigma}$. Therefore every connected static 
Galilei transformation is the fourth power of a coset transformation.   

As is well-known, according to Noether's theorem, there exists a conserved 
quantity corresponding to every continuous symmetry.  The conserved quantities 
for the usual Galilean transformations are standard and will not be repeated
here. The conserved quantities for the $SL(2,R)$ symmetry can be derived as 
follows: The invariance of the Action implies
$$\delta L (x, \dot x, t) = {d {\cal F}\over d t}\eqno(30)$$
For time-independent Lagrangians $L(x_i,\dot x_i)$, we have
$$\delta L = {\partial L\over\partial x_i} \delta x_i + {\partial L\over
\partial\dot x_i} \delta\dot x_i = {d\over d t} \left( {\partial L\over\partial
\dot x_i} \delta x_i \right)\eqno(31) $$
where the Euler-Lagrange equations have been used in the second equality.  
Combining the two expressions for $\delta L$, we get a conservation law
$$ {d\over d t} \left( {\partial L\over\partial\dot x_i} \delta x_i - {\cal F}
\right) = 0\eqno(32) $$
The $SL(2,R)$ transformations are 
$$\eqalign{t &\to \tau = {\alpha t+\beta\over \gamma t+\delta} \cr
x(t) &\to \xi(\tau) = {x(t)\over \gamma t+\delta} = (\alpha-\gamma\tau) 
\, x\left({\delta\tau-\beta\over-\gamma\tau+\alpha}\right)}\eqno(33) $$ 
For infinitesimal transformations, $\alpha=1+\varepsilon$ and
$\delta=1-\varepsilon$ (to ensure $\alpha\delta-\beta\gamma=1$) with
infinitesimal $\beta$, $\gamma$, $\varepsilon$,
$$ \eqalign{\delta t &= \beta + 2\varepsilon t - \gamma t^2 \cr
\delta x(t) &= (\varepsilon-\gamma t)\, x(t)
- (\beta + 2\varepsilon t-\gamma t^2)\, \dot x(t)}\eqno(34) $$
The change of $L=(m/2) \dot x^2$ is given by
$$ \delta L = m \dot x \cdot \delta\dot x 
= m \dot x \cdot \bigl( -\gamma x - (\varepsilon-\gamma t)\dot x
- (\beta + 2\varepsilon t-\gamma t^2) \ddot x \bigr)
\eqno(35) $$
It is easily seen to be the total time derivative of 
$$ {\cal F} = - \bigl( \beta + 2\varepsilon t-\gamma t^2 ) {m\dot x^2\over 2}
- \gamma {m x^2\over2}\eqno(36) $$
Therefore, we obtain the conserved quantity
$$ X = m\dot x\cdot\delta x - {\cal F} 
=- \bigl( \beta + 2\varepsilon t-\gamma t^2 ) {m\dot x^2\over 2}
+ (\varepsilon-\gamma t) m x \cdot \dot x
+ \gamma {m x^2\over2}\eqno(37) $$
Extracting the coefficients of $\beta,~\varepsilon$ and $\gamma$ we get 
$$ \eqalign{X &= - \beta H + \varepsilon D + \gamma A \cr
H &= {p^2\over2m}\cr
D &= p\cdot \left( x - {t p\over m} \right)\cr
A &= {(t p-m x)^2\over2m}  }\eqno(38)$$
where $p = m\dot x$ and $H,~D,~A$ are the conserved quantities related to time 
translations, dilatations, and expansions respectively. The following 
interesting observation about the conserved quantities can now be made: The 
Noether's theorem can also be used to show that the conserved quantities 
corresponding to the usual translation and boost symmetries are $p_i = m\dot 
x_i$ and $K_i = tp_i - mx_i$ respectively; hence it follows that $A = K^2/2m$ 
is related to $K$ in the same way as the Hamiltonian is related to the momenta.
Mathematically, $H$, $A$ and $D = -p\cdot K/m$ form the adjoint representation 
of the $SL(2, R)$, while $p$ and $K$ transform as a doublet, and rotations 
are invariant. This concludes the discussion of the symmetries of the classical
non-relativistic point particle.

We shall now consider the quantum mechanical generalisation of the above 
results. For this it is convenient to think of the wavefunction of the particle
as a non-relativistic field. For the usual ten-parameter Galilei group $G$, it 
is well-known that, in the field theoretic realisation, there is a 
one-parameter mass group $M$ that commutes with $G$. This is called the 
central extension and has the effect of modifying the Lie algebra of $G$ in a 
non-trivial manner, while at the same time preserving the Jacobi identity. It 
turns out that a similar feature holds for the twelve-parameter group 
${\cal G}$. 

In the field representation the conserved quantities of ${\cal G}$ are 
$${\bf P} = -i\nabla,~~{\bf J} = -i{\bf x}\times\nabla~~~{\bf K} = -it\nabla
- m{\bf x}\eqno(39a)$$
$$D = i(2t{\partial\over\partial t} + {\bf x}\cdot\nabla + {3\over 2}),~~~A = 
-i(t^2{\partial\over\partial t} -{m\over 2}{\bf x}^2 + t{\bf x}\cdot\nabla + 
{3\over 2}t),~~~ H = i {\partial\over\partial t}\eqno(39b)$$
where the factors of $3/2$ appear because we have Weyl ordered products of 
position and momentum in the quantum theory and set $\hbar = 1$. 
${\bf P,~J,~K}$ generate translations, rotations and boosts respectively, which 
constitute the connected static Galilei group (21). $H$, $D$, and $A$  
produce time translations, dilations, and {\it expansions} which together  
constitute the $SL(2, R)$ group. The central extension of the standard Lie 
algebra of the former
$$[J_i, J_k] = i\epsilon_{ikl} J_l,~[J_i, P_k] = i\epsilon_{ikl} P_l,~ 
~[P_i, P_k] = 0$$
$$[K_i, K_k]= 0,~[J_i, K_k] = i\epsilon_{ikr}K_r,~[K_i, P_k] = -(im\delta_{ik})
$$
$$[P_i, H] = 0 ~~~[K_i, H] = -iP_i,~~~[J_i, H] = 0$$
is augmented by the additional relations involving the $SL(2, R)$ generators 
$$[D, J_i] = 0,~~~[D,  K_i] = iK_i,~~~[D, P_i] = -iP_i$$
$$[A, J_i] = 0,~~~[A, K_i] = 0,~~~[A, P_i] = i K_i$$
$$[D, H] = -2iH,~~~[A, H] = iD,~~~[D, A] = 2iA$$
to give the Lie algebra of the full group ${\cal G}$. The bracketed term in the
$[K_i, P_j]$ commutator is the non-trivial modification that the central
extension brings about in the Lie algebra with $m$ -- to be physically 
identified with the mass of the particle -- standing for the value that 
the generator of $M$ takes in the given representation.     

Since the existence of the central extension implies that  
$$[K_i, P_j] = -(im\delta_{ij})\eqno(40)$$ 
the groups of translations $T({\bf a}) = e^{i{\bf a}\cdot {\bf P}}$ and boosts 
$B({\bf v}) = e^{i{\bf v} \cdot{\bf K}}$ no longer form a direct product, but 
a Heisenberg-Weyl group defined by the relation
$$T({\bf a})B({\bf v}) = B({\bf v})T({\bf a})e^{im{\bf  a}\cdot{\bf v}}
\eqno(41)$$ 
The action (24) of $SL(2,R)$ on ${\bf a}$ and ${\bf v}$ induces the 
transformations
$$\sigma^{-1} \pmatrix{{\bf K}\cr{\bf P}} \sigma = \pmatrix{\alpha&\beta\cr 
\gamma&\delta} \pmatrix{{\bf K}\cr{\bf P}}\eqno(42) $$ 
of the generators ${\bf K}$ and ${\bf P}$. The commutator (40) does not change 
under these transformations because of the condition 
$\alpha\delta-\beta\gamma=1$.  Thus the central extension is compatible with 
$SL(2,R)$ and the full invariance group of the quantised system is 
the central extension of ${\cal G}$.

Indeed it was in the quantum theory that the group ${\cal G}$ was first 
discovered by Niederer who showed that it is the maximal kinematical invariance
group of the free particle Schr\" odinger equation [4]. Since the invariance 
under the Galilei group is well-known, it suffices to verify that the 
Schr\"odinger equation    
$$i {\partial\psi\over \partial t} + {1\over 2m}{\partial^2\psi
\over\partial x_i^2} = 0\eqno(43)$$ 
remains form-invariant under $SL(2, R)$. It is easily checked that this
is accomplished by the following transformation of the wavefunction that 
is generated by $A,~H,~{\hbox{and}}~D$ of the $SL(2,R)$ group. 
$$\psi ({\bf x}, t)~~ \propto~~ (\gamma\tau - \alpha )^{3\over 2} e^{i{F\over 
2}} \psi({\bf \xi} , \tau) \eqno(44)$$
where $F$ is determined through (8) and (9).  

As promised, we now sketch a road map to an astrophysical application of the 
results of this paper. As is explained in detail in [2], recent experimental
programmes  try to simulate astrophysical systems like supernova explosions
in the laboratory by creating implosions in inertial confinement plasmas. 
This research is inspired by a remarkable similarity which was observed in the 
results of numerical models of 1987A supernova observations and results of 
numerical simulations of experiments involving plasma implosions. Referring 
the reader to [2] for further details of this research programme, we note that 
this is a puzzling observation because the former system involves very large 
length and time scales whereas the latter involves very small scales. A 
theoretical explanation of this intriguing similarity was given in [1] and can 
be traced to the symmetry properties of the fluid dynamic equations [5, 6] 
that describe both stellar structure and the plasma state. As conjectured in 
[1], this symmetry has its origin in the symmetries of the free point particle. 
In particular the symmetry responsible for mapping supernova explosions to 
plasma implosions is the fluid mechanical analogue of the {\it expansion} 
transformations discussed in this paper. To make this connection more precise, 
we note that the analysis of the single free point particle can be carried over
to the non-interacting many particle case in a straightforward manner. Indeed, 
using the expression for the Hamiltonian of an ensemble of free point particles
labelled by $I$, and the expression for the momenta ${\bf p}_{~\atop I}$, 
$$H = \sum_I {{\bf p}_{~\atop I}^2\over 2m}~~~{\hbox{and}}~~~{\bf p}_{~\atop I}
 = m\dot{\bf x}_{~\atop I} \eqno(45)$$
it is easy to see that the corresponding Liouville equation stating the 
invariance of the density of particles $\rho$ along the flow  
$$ {d\rho\over dt} = {\partial\rho\over \partial t} + \{\rho , H\}~~~= 0
\eqno(46)$$ 
also has a maximal invariance group given by (26). As is well-known, the 
Liouville equation can be converted into the Boltzmann equation by 
expressing all the momenta in terms of the velocities. Therefore, the 
symmetry group carries over to the collisionless Boltzmann equation. 
Further, in the continuum limit, one can use the standard procedure of 
deriving the various fluid dynamic equations as moments of the Boltzmann 
equation[7]. The simplest example is the set of Euler equations [8]
$$ D\rho = -\rho\nabla\cdot {\bf u} \eqno(47a)$$
$$ \rho D{\bf u} = -\nabla p \eqno(47b)$$
$$ D\epsilon = -(\epsilon + p)\nabla\cdot {\bf u}\eqno(47c)$$
where $\rho$, {\bf u}, $p$ and $\epsilon$ stand for the density, the velocity 
vector field, the pressure, and the energy density of the fluid respectively. 
The convective derivative $D$ in the above equations is defined by $D = 
{\partial\over\partial t} + {\bf u}\cdot\nabla$. The above differential 
equations of fluid flow are usually augmented by an algebraic condition called 
the polytropic equation of state which relates the pressure to the energy 
density thus:
$$p = (\gamma_o - 1)\epsilon\eqno(47d)$$
where $\gamma_o$ is a constant called the polytropic exponent. For the 
ensemble of free non-relativistic point particles being considered here, 
$\gamma_0$ takes the value 5/3.  
The maximal symmetry group ${\cal G}$ therefore extends to the fluid dynamic
equations which explains the observed similarity between supernova explosions
and plasma implosions.  

Although we have considered free particles so far, it is interesting to note
that the symmetry group discussed in this paper extends to an interesting
class of interacting many-body problems namely, those for which the potential
is an inverse square of the coordinate differences. In one dimension these 
include the so-called Calogero-Moser models [9] with Hamiltonian 
$$ H = {1\over 2}\Bigl(\sum_{I=1}^N p_{~\atop I}^2 + g^2\sum_{I\neq J}{1\over 
(x_{~\atop I} - x_{~\atop J})^2}\Bigr)\eqno(48)$$
These models are integrable and admit {\it exclusion statistics} -- an 
exciting area of current research [10]. In two dimensions it may be shown that 
models with Hamiltonians of the form 
$$H = {1\over 2}\Bigl(\sum_{I=1}^N \bigl({\bf p}_{~\atop I} - e{\bf A}
({\bf x}_{~\atop I})\bigr)^2\Bigr),~~~
{\hbox{where}}~~~A_k({\bf x}_{~\atop I})\propto\sum_{J\neq I} 
{\epsilon_{kl} ({\bf x}_{~\atop I} - {\bf x}_{~\atop J})_{~\atop l}
\over \mid x_{~\atop I} - x_{~\atop J}\mid^2}\eqno(49)$$
which describe a gas of anyons -- particles having arbitrary spin and 
statistics -- have potentials which admit the symmetry group ${\cal G}$. As is 
well-known, anyons are of interest because they appear as excitations in 
fractional quantum Hall systems [11]. 
 
To conclude, we have investigated the maximal kinematical invariance group of 
a free nonrelativistic point particle and have found that it is bigger than
the Galilei group. It is  a semi-direct product of the form $SL(2,R)
\wedge G$ where $G$ is the static Galilei group. As shown in this paper, 
this group is in fact the maximal invariance group of a host of interesting 
systems in which the physics content is captured by the quintessential free 
nonrelativistic point particle. However there exists a class of interacting 
many particle systems -- of which the well-known Calogero-Moser models and the 
anyon model are of particular interest -- for which this is also true. 
\vfil\eject
\centerline {\bf References}
\bigskip
\item {1. } L. O'C Drury and J. T. Mendon\c ca, ``Explosion implosion duality
and the laboratory simulation of astrophysical systems,'' Physics of Plasmas,
{\bf 7} No.12 (2000) 5148-5142. 
\item {2. } Supernova Hydrodynamics Up Close: Science and Technology Review,
Jan'/Feb' 2000; http://www.llnl.gov/str;  I. Hachisu {\it et al.},
``Rayleigh-Taylor instabilities and mixing in the helium star models for Type
Ib/Ic supernovae,''
Astrophysical Journal {\bf 368} (1991) L27--30; 
H. Sakagami and K. Nishihara, ``Rayleigh-Taylor instability on the
pusher-fuel contact surface of stagnating targets,'' 
Physics of Fluids {\bf B 2} (1990) 2715--2730.
\item {3. } P. J. Olver, Applications of Lie Groups to Differential 
Equations; Springer-Verlag (1986). 
\item {4. } U. Niederer, ``The maximal kinematical invariance group of the free Schr\" odinger equation,'' Helvetica Physica Acta, {\bf 45} (1972) 802--810. 
\item {5. } L. O'Raifeartaigh and V. V. Sreedhar, ``The maximal kinematical
invariance group of fluid dynamics and explosion-implosion duality,''
hep-th/0007199. 
\item {6. } M. Hassaine and P. A. Horvathy, ``Field-dependent symmetries of a 
non-relativistic fluid model,'' math-ph/9904022, ``Symmetries of fluid 
dynamics with polytropic exponent,'' Phys. Lett. {\bf A279} (2001) 215-222;
S. Deser, R. Jackiw and A. P. Polychronakos, ``Clebsch (string)
parameterization of 3-vectors and their actions,'' physics/0006056; 
R. Jackiw, V. P. Nair and So-Young Pi, ``Chern-Simons reduction and
non-Abelian fluid mechanics,'' Physical Review {\bf D 62} (2000) 085018, 
hep-th/0004084; 
R. Jackiw, ``A particle field theorist's lectures on supersymmetric, 
non-Abelian fluid mechanics and D-branes,'' physics/0010042. 
\item {7. } K. Huang, Statistical Mechanics; John Wiley and Sons (1967). 
\item {8. } L. D. Landau and E. M. Lifshitz, Fluid Mechanics, Pergamon Press
(1959).
\item {9. } F. Calogero, ``Exactly solvable one-dimensional many body 
problems,'' Lett. Nuov. Cim. {\bf 13} (1975) 411; 
J. Moser, ``Three integrable Hamiltonian systems connected with isospectral
deformations,'' Adv. Math.  {\bf 16} (1975) 197--220. 
\item {10. } A. Polychronakos, ``Generalized statistics in one dimension,''
hep-th/9902157; Lectures at the Les Houches Summer School in Theoretical 
Physics, Session 69: Topological Aspects of Low-dimensional Systems.  
\item {11. } F. Wilczek, Fractional Statistics and Anyon Superconductivity;
World Scientific (1990).

\bye